\documentstyle[11pt]{article}
\def\ni{\noindent}

\def\abstract#1{\vskip 15pt\midinsert\narrower\narrower\smallskip
   \ctr{ABSTRACT}\vskip5pt\noindent{#1}\smallskip\endinsert}

\newcommand{\beq}{\begin{equation}}
\newcommand{\eeq}{\end{equation}}
\newcommand{\bqn}{\begin{eqnarray}}
\newcommand{\eqn}{\end{eqnarray}}
\newcommand{\bqns}{\begin{eqnarray*}}
\newcommand{\eqns}{\end{eqnarray*}}
\newcommand{\bary}{\begin{array}}
\newcommand{\eary}{\end{array}}
\newcommand{\non}{\nonumber}

\newcommand{\dt}{\Delta}
\newcommand{\ot}{\otimes}

\newcommand{\ra}{\rightarrow}
\newcommand{\ep}{\epsilon}
\newcommand{\vep}{\varepsilon}
\newcommand{\ga}{\gamma}

\newcommand{\h}{\frac{1}{2}}
\newcommand{\hs}{\hspace}
\newcommand{\sg}{\sigma}

\newtheorem{defn}{Definition}[section]
\newtheorem{prop}{Proposition}[section]
\begin{document} {\flushright (July, 1999)} \vspace{44pt}
\begin{center}

{\Large\sc{\bf Colored solutions of Yang-Baxter equation from
representations of $U_{q}gl(2)$}}

\baselineskip=12pt \vspace{35pt}

\renewcommand{\thefootnote}{\fnsymbol{footnote}} Pi-Gang
Luan$^{1}$,\footnote[1]{\it E-mail address:
lpg@sansan.phy.ncu.edu.tw} H. C. Lee$^{1,2}$\footnote[2]{\it
E-mail address: hclee@sansan.phy.ncu.edu.tw} and R. B.
Zhang$^{3}$\footnote[3]{\it E-mail address:
rzhang@maths.uq.edu.au}

\medskip

{\it $^1$Department of Physics and Center for Complex Systems,
National Central University, Chung-li, Taiwan, ROC\\}

{\it $^{2}$National Center for Theoretical Sciences, P.O. Box
2-131, Hsinchu, Taiwan, ROC\\and\\}

{\it $^{3}$Department of Mathematics, University of Queensland,
Brisbane, QLD 4072, AUSTRALIA}

\vspace{15pt}

\end{center}


\narrower{\ni{\bf Abstract}. We study the Hopf algebra structure
and the highest weight representation of a multiparameter version
of $U_{q}gl(2)$. The commutation relations as well as other Hopf
algebra maps are explicitly given. We show that the multiparameter
universal ${\cal R}$ matrix can be constructed directly as a
quantum double intertwiner, without using Reshetikhin's
transformation. An interesting feature automatically appears in
the representation theory: it can be divided into two types, one
for generic $q$, the other for $q$ being a root of unity. When
applying the representation theory to the multiparameter universal
${\cal R}$ matrix, the so called standard and nonstandard colored
solutions $R(\mu,\nu; {\mu}', {\nu}')$ of the Yang-Baxter equation
is obtained.

\vspace{10pt}
\vskip 0.7 in

\baselineskip=12pt
\section{Introduction}
As is well known, the Yang-Baxter equation (YBE) \cite{Bxt,Ma}
plays an essential role in the study of quantum groups (QG) and
quantum algebras (QA) \cite{Df,Jb,FRT,Tt,Tt1,Tjin,Ma}, integrable
models \cite{LFK,KSk,MN,PS}, as well as in the construction of
knot or link invariants \cite{Kf,Wu,Lee,AD,Lee2,Lee1,GLW}. For
instance, in the Faddeev--Reshetikhin--Takhtajan (FRT) approach
\cite{FRT,Tt,Tt1} to construct quantum groups or quantum algebras,
one has to find an $R$ matrix, which is a matrix solution of YBE
\cite{Ma}, then using this $R$ matrix as the input, substituting
it into the $RTT$ or $RLL$ relations to get the quantum group or
quantum algebra as the output.

There are various methods to find the appropriate $R$ matrix. One
way is to borrow an $R(u)$ matrix from the integrable model
\cite{Ma} and then taking appropriate limit to remove the spectral
parameter $u$. The second method is to solve the matrix YBE
directly \cite{Lee1,CLS,GLX}. In this approach one usually assumes
an $R$ with prescribed nonzero elements, and impose some
restrictions on them to find a class of solutions. Some $R$
matrices obtained in this way have unexpected interesting features
\cite{Lee,GLW,GX,GOW}, so a number of authors call them
``nonstandard'' solutions \cite{JGW,GWu,GWu1}.

Many known quantum algebras belong to the category of
quasitriangular Hopf algebras (QTHA) \cite{Tjin,Ma}. This
observation provides us an alternative approach to find the $R$
matrix\cite{KBM,BH1}. When applying representation theory to the
universal ${\cal R}$ matrix\footnote[4]{We denote the universal
algebraic solution of YBE by ${\cal R}$ and the matrix solution by
$R$} \cite{Tjin,Ma} of a QTHA, the desired $R$ matrix is obtained.
To get more interesting solutions, people also try different
methods to add parameters that appearing in the $R$ matrix . This
cause the development of multiparameter deformations
\cite{Rs,Jing,Sc,SWZ,BH,KBM,CJ,BH1} Hopf algebras and $q$-boson
realizations \cite{GLS,GLS1,GF,GSX,GWa} with $q$ being a root of
unity. These solutions are sometimes called ``colored'' solutions
\cite{GLW,GOW,KBM}. Although the $q$-boson realization method is
very powerful in constructing representations of quantum groups or
quantum algebras, it's hardly to manifest the Hopf algebra
structures.

In this paper we study $U_{q}gl(2)$. We show that due to the
commuting element $J$, it is possible to introduce additional
parameters $t$, $u$ and $v$, and hence gives us a multiparameter
version of Hopf maps and multiparameter universal ${\cal R}$
matrix. We then explain how to get the same ${\cal R}$ from
quantum double constructions. In this way the Hopf algebra
structure is preserved and emphasized. For the representations of
$U_{q}gl(2)$, we only consider the highest weight representations.
Under the finite dimension restriction, two categories of
representation appears automatically. When applying this
representation theory to ${\cal R}$, the standard and nonstandard
colored solutions are obtained and are consistent with
literature's results.

This paper is organized as follows: In section 2, we review some
basic definitions and properties of Hopf algebras, quasitriangular
Hopf algebras and quantum double. In section 3, different
selections of universal ${\cal R}$ matrices are given, and
compared to the result obtained from Reshetikhin's transformation
\cite{Rs}. In section 4, the highest weight representations are
studied and applied to ${\cal R}$ to obtain matrix solutions $R$.
In section 5, colored solutions are obtained and compared to the
literature's results. Section 6 is devoted to concluding remarks.
\section{Hopf algebras, quasi-triangular Hopf algebras and Quantum double}

In this section we give brief review of some definitions and
properties of Hopf algebras (HA) and quasi-triangular Hopf
algebras (QTHA), as well as their relations to the notion of
quantum double (QD) \cite{Tjin,Ma}. These ideas will then be used
in our latter discussions of the multiparameter $U_{q}gl(2)$.\\

\ni{\bf A.\/} {\bf Hopf algebras\/}\\

A Hopf algebra is an associative algebra $A$ with five basic maps(
in this paper, we call them Hopf maps), they are four
homomorphisms: $m:A\ot A\ra A$ ($multiplication$), $\dt:A\ra A\ot
A$ ($coproduct$), $\eta:C\ra A$ ($inclusion$), $\vep:A\ra C$
($counit$) and one antihomomorphism: $S:A\ra A$ ($antipode$). They
satisfy the following relations for any $a\in A$: \beq \bary{l}
(\dt \ot id)\dt(a) =(id \ot \dt)\dt(a)\\ \mbox{}\\ (\vep\ot
 id)\dt(a)=(id\ot\vep)\dt(a)
=id(a)=a\\ \mbox{}\\ m(S\ot id)\dt(a)=m(id\ot S)\dt(a) =\eta \circ
\vep(a)=\vep(a)1 \eary\label{ha} \eeq where $id$ represents the
$identity$ $map$. To be more precise, we use the notation
$(A,m,\dt$,$\eta,\vep,S)$ instead of $A$ to denote a Hopf algebra.
With these ideas in mind, the following propsition will be
apparent:
\begin{prop}
Replacing $\dt$ by $\bar{\dt}=\dt^{'}$ and $S$ by
$\bar{S}=S^{-1}$, the algebra $(A,m,\bar{\dt},\eta,\vep,\bar{S})$
is also a Hopf algebra.
\end{prop}
Here $\dt^{'}$ denotes the opposite coproduct, which maps any
$a\in A$ to $A\ot A $ as: \beq
\dt^{'}(a)=\sg\circ\dt(a)=\sum_{i}c_{i}\ot b_{i}\;\;\mbox{if}\;\;
\dt(a)=\sum_{i}b_{i}\ot c_{i} \eeq and $S^{-1}$ is defined as the
inverse of $S$: \beq S(S^{-1}(a))=S^{-1}(S(a))=a \eeq\\

\ni{\bf B.} {\bf Quasitriangular Hopf algebras\/}\\

A Quasitriangular Hopf algebra (QTHA)is a Hopf algrbra equipped
with an element ${\cal R}\in A\ot A$ which is the solution of the
algebraic version of YBE. We start with the definition.
\begin{defn}
Let ${\cal A}$=$(A,m,\dt,\eta,\ep,S)$ be a Hopf algebra and ${\cal
R}$ ($intertwiner$) an invertible element in $A\ot A$, then the
pair$({\cal A},{\cal R})$ is called a QTHA if for any $a\in A$ we
have
\begin{description}
\item[(i)] ${\cal R}\dt(a)=\dt^{'}(a){\cal R}$
\item[(ii)] $(\dt \ot id){\cal R}={\cal R}_{13}{\cal R}_{23}$
\item[(iii)] $(id\ot\dt){\cal R}={\cal R}_{13}{\cal R}_{12}$
\end{description}
\end{defn}
In addition, three further relations are satisfied: \beq \bary{l}
{\cal R}_{12}{\cal R}_{13}{\cal R}_{23}={\cal R}_{23}{\cal
R}_{13}{\cal R}_{12},\\ \mbox{}\\ (S\ot id){\cal R}=(id\ot
S^{-1}){\cal R} ={\cal R}^{-1},\\ \mbox{}\\ (\vep\ot id){\cal R}
=(id\ot \vep){\cal R}=1, \eary \eeq The first line is the
Yang-Baxter equation.

As in the case of Hopf algebras, we denote $(A, {\cal R }, m, \dt,
\eta, \vep, S)$ as a QTHA. From (i) of definition {\bf 2.1}, we
immediately find
\[ \bary{lll}
{\cal R}\dt(a)=\dt^{'}(a){\cal R},& \mbox{} & (\sg\circ{\cal
R})\dt^{'}(a)=\dt(a)(\sg\circ{\cal R}),\\ \mbox{}&
\mbox{}&\mbox{}\\ {\cal R}^{-1}\dt^{'}(a)=\dt(a){\cal
R}^{-1},&\mbox{}&(\sg\circ{\cal R}^{-1})\dt(a)
=\dt^{'}(a)(\sg\circ{\cal R}^{-1}). \eary \] \ni Define ${\cal
R}^{(+)}=\sg\circ{\cal R}$,
 ${\cal R}^{(-)}={\cal R}^{-1}$ and
 $\bar{\cal R}=\sg\circ{\cal R}^{-1}$,
and denote $\dt^{'}$ as $\bar{\dt}$, then \beq \bary{lll} {\cal
R}\dt=\dt^{'}{\cal R},&\mbox{}&\bar{\cal R}\dt=\dt^{'}\bar{\cal
R},\\ \mbox{}&\mbox{}&\mbox{}\\ {\cal
R}^{(+)}\bar{\dt}=\bar{\dt^{'}}{\cal R}^{(+)},& \mbox{}&{\cal
R}^{(-)}\bar{\dt}=\bar{\dt^{'}}{\cal R}^{(-)}. \eary \eeq

These observations lead to the following result:\\
\begin{prop}
If $(A, {\cal R}, m, \dt, \vep, S, \eta)$ is a QTHA, then $(A,
\bar{\cal R}, m, \dt, \vep, S, \eta)$, $(A, {\cal R}^{(+)}$,$ m$,
$\dt^{'}$, $\vep$, $S^{-1}$, $\eta)$ and $(A, {\cal R}^{(-)}, m,
\dt^{'}, \vep, S^{-1}, \eta)$ are all QTHAs.
\end{prop}

It can be easily proved by using the definition {\bf 2.1} and
equation (\ref{ha}). This theorem tells us that for a pair
$(\dt\,,\,S)$ , there are two universal ${\cal R}$ matrices:
${\cal R}$ and $\bar{\cal R}=\sg\circ {\cal R}^{-1}$, both can be
used as intertwiner in a QTHA. Now let's turn to the discussion of
$quantum$ $double$\cite{Tjin,Ma} .\\

\ni{\bf C.} {\bf Quantum double\/}\\

Suppose we have a Hopf algebra $A$, which spanned by basis
$\{e_{i}\}$. By introducing a nondegenerate bilinear form
$\langle\;,\;\rangle$, we can define $A$'s dual algebra $A^{o}$,
which spanned by $\{e^{i}\}$; here $\langle e^{i}\,,\,e_{j}
\rangle=\delta^{i}_{j}$. Then all the Hopf maps of $A^{o}$ can be
defined in terms of $\langle\;,\;\rangle$. Introducing the
$intertwiner$: \beq {\cal R}=\sum_{i}e_{i}\ot e^{i}, \eeq then the
commutation relations between $A$ and $A^{o}$ can be established
via the relation
\[
{\cal R}\dt(a)=\dt^{'}(a){\cal R},\;\;\;\mbox{for}\;\; a\in A
\;\;\mbox{ or} \;\; A^{o},
\]
which tells us how to expand an $e^{i}e_{j}$ type product as the
sum of $e_{i}e^{j}$ type products. Choosing $\{e_{i}e^{j}\}$ as
basis, one can ``combine'' $A$ and $A^{o}$ to form an enlarged
algebra $D(A)$--the $quantum$ $double$ of $A$, and treat $A$ or
$A^{o}$ as its subalgebra. Then $D(A)$ can be proved to be a QTHA
equipped with ${\cal R}=\sum_{i}e_{i} {\ot} e^{i}$ as its
intertwiner( universal ${\cal R}$ matrix ). In other words, a QTHA
is a quantum double of its subalgebra. In the next section, we
will show that the $U_{q}gl(2)$ is indeed a quantum double as well
as a QTHA.
\section{Universal ${\cal R}$ matrix of $U_{q}gl(2)$}

We define Our version of $U_{q}gl(2)$ algebra as a multiparameter
QTHA generated by $(H,J,X^{+},X^{-})$ with the commutation
relations
\[
[J\,,\,H]=[J\,,\,X^{\pm}]=0,
\]
\beq [H\,,\,X^{\pm}]=\pm 2 X^{\pm},\hs{5mm}\label{cmt} \eeq
\[
[X^{+},\,X^{-}]=\frac{q^{H}t^{-J}-q^{-H}t^{J}}{q - q^{-1}},
\]
and additional Hopf maps: \beq\bary{ll} \mbox{}&\dt(H)=H\ot 1 +
1\ot H,\;\;\;\;\;\dt(J)=J\ot 1+1\ot J,\\ \mbox{}&\mbox{}\\
\mbox{coproduct:}\;\;\;&
 \dt(X^{+})=q^{-\h H}(utv)^{\h J}\ot X^{+}+X^{+}\ot q^{\h
H}(utv^{-1})^{-\h J},
\\
\mbox{}&\mbox{}\\ \mbox{}& \dt(X^{-})=q^{-\h H}(u^{-1}tv^{-1})^{\h
J}\ot X^{-}+X^{-}\ot q^{\h H}(u^{-1}tv)^{-\h J},
\eary\label{ddd}\eeq \beq \mbox{antipode:}\;\;\;\;\;\;
S(H)=-H,\;\;S(J)=-J,\;\; S(X^{\pm})=-q^{\pm 1}v^{\mp
J}X^{\pm},\hs{13mm}\label{SSS} \eeq \beq
\mbox{counit:}\;\;\;\;\;\;
\vep(H)=\vep(J)=\vep(X^{\pm})=0.\hs{52mm}\label{eee} \eeq The
universal ${\cal R}$ matrix is defined by \beq {\cal R}= {\cal
R}_{0}\sum^{\infty}_{n=0} \frac{(1-q^2)^{n}}{\{n\}_{q^2}!}
q^{n(n-1)}q^{\frac{n}{2} (H\ot 1-1\ot H)} ((uvt^{-1})^{\h
J}X^{-})^{n}\ot ((uv^{-1}t)^{\h J}X^{+})^{n}, \label{Rv}\eeq where
\beq {\cal R}_{0}= q^{-\h H\ot H}t^{\h (H\ot J+J\ot H)}u^{\h (H\ot
J-J\ot H)},\label{R0} \eeq $t$, $u$ and $v$ are arbitrary
parameters and $\{n\}_{q^2}$, $\{n\}_{q^2}!$ are defined by
 \beq\bary{c}
\{n\}_{q^2}=\frac{\displaystyle 1-q^{2n}}{\displaystyle
1-q^2}=q^{n-1}[n]_{q},\\ \mbox{}\\
\{n\}_{q^{2}}!=\displaystyle{\prod^{n}_{j=1}} \{j\}_{q^{2}}= q^{\h
n(n-1)}[n]_{q}!, \eary\eeq with $\{0\}_{q^2}!=[0]_{q}!=1$. Note
that the commuting element $J$ appearing in this algebra causes
the expression of Hopf maps has many different choices. For
example, the parameter $t$ in the last commutation relation of
(\ref{cmt}) is not essential. One can always absorb the factor
$t^{-J}$ into $q^{H}$ by defining $q^{H}t^{-J}=q^{H'}$ and rename
$H'$ by $H$. However, in order to reflect the fact that $J$ can be
arbitrarily `mixed' with $H$, in this paper we shall always retain
the parameter $t$. On another hand, two arbitrary parameters $u$
and $v$ are allowed to appearing in the definitions of $\dt$ and
$S$, although they do not explicitly appear in (\ref{cmt}).

Note that under the transformation: \beq \tilde{X^{\pm}}=v^{\mp \h
J} X^{\pm},\label{til} \eeq the commutation relations (\ref{cmt})
 will not change its form.
Moreover, \ref{til} simplifies the form of $\dt$ and $S$ on
$X^{\pm}$: \beq \dt(\tilde{X^{\pm}})=(q^{-\h H}t^{\h J})u^{\pm \h
J}\ot \tilde{X^{\pm}}+\tilde{X^{\pm}}\ot (q^{\h H}t^{-\h J})u^{\mp
\h J}\label{til1}, \eeq \beq S(\tilde{X^{\pm}})=-q^{\h
H}\tilde{X^{\pm}} q^{-\h H}=-q \tilde{X^{\pm}}.\label{til2} \eeq
Furthermore, the universal ${\cal R}$ matrix now becomes: \beq
{\cal R}={\cal
R}_{0}\sum^{\infty}_{n=0}\frac{(1-q^2)^{n}}{[n]_{q}!} q^{-\h
n(n-1)}((ut^{-1})^{\h J}q^{\h H}\tilde{X^{-}})^{n}\ot ((ut)^{\h
J}q^{-\h H}\tilde{X^{+}})^{n}. \label{R} \eeq In the following, we
shall use $\tilde{X^{\pm}}$ as generators.

As stated in the last section,corresponding to the same pair
$(\dt\,,\,S)$, there is another universal ${\cal R}$ matrix: \beq
\bar{\cal R}={\bar{\cal
R}}_{0}\sum^{\infty}_{n=0}\frac{(1-q^{-2})^{n}}{[n]_{q}!} q^{\h
n(n-1)} ((ut)^{-\h J}q^{\h H }\tilde{X^{+}})^{n}\ot ((u^{-1}t)^{\h
J}q^{-\h H }\tilde{X^{-}})^{n} \eeq with \beq{\bar{\cal
R}}_{0}=\sg\circ{\cal R}^{-1}_{0}=q^{\h H\ot H}t^{-\h (H\ot J+J\ot
H)}u^{\h (H\ot J-J\ot H)}. \eeq

Similarly, if we use $\bar{\dt}={\dt}'$ and $\bar{S}=S^{-1}$ as
$coproduct$ and $antipode$ respectively then, for the pair
$(\bar{\dt}\,,\,\bar{S})$, we have the following two universal
${\cal R}$ matrices: \bqn {\cal R}^{(+)} &=&{\cal
R}^{(+)}_{0}\sum^{\infty}_{n=0}\frac{(1-q^2)^{n}}{[n]_{q}!}q^{-\h
n(n-1)}((ut)^{\h J}q^{-\h H}\tilde{X^{+}})^{n}\ot ((ut^{-1})^{\h
J}q^{\h H}\tilde{X^{-}})^{n},\non\\ {\cal R}^{(+)}_{0}
&=&\sg\circ{\cal R}_{0}=q^{-\h H\ot H}t^{\h (H\ot J +J\ot
H)}u^{-\h (H\ot J-J\ot H)}, \eqn and \bqn {\cal R}^{(-)}&=&{\cal
R}^{(-)}_{0}\sum^{\infty}_{n=0}\frac{(1-q^{-2})^{n}}{[n]_{q}!}q^{\h
n(n-1) }((u^{-1}t)^{\h J})q^{-\h H }\tilde{X^{-}})^{n}\ot
((ut)^{-\h J}q^{\h H}\tilde{X^{+}})^{n},\non\\ {\cal
R}^{(-)}_{0}&=&{\cal R}^{-1}_{0}=q^{\h H\ot H}t^{-\h (H\ot J+J\ot
H)}u^{-\h (H\ot J-J\ot H)}. \eqn

These universal ${\cal R}$ matrices can be compared to the
literature \cite{BH}-\cite{BH1}. However, since different authors
adopt different conventions in the definition of $\dt$ and $S$,
thus we have to properly choose one ${\cal R}$ from the set
$\{{\cal R}$, $\bar{\cal R}$, ${\cal R}^{(+)}$, ${\cal
R}^{(-)}\}$. Note that the parameters $t$, $u$ and $v$ can be
freely choosen due to the fact that there exist the commuting
generator $J$. If we use $H_1=H-\alpha_1 J$ and $H_2=H-\alpha_2 J
$ as generators instead of $H$ and $J$ where
$q^{\alpha_1}=u^{-1}t$ and $q^{\alpha_2}=ut$, then the universal
${\cal R}$ matrix can be expressed as the following simple form (
here we drop a trivial commuting factor $q^{-\h
\alpha_{1}\alpha_{2}J\ot J}$ ): \beq {\cal R} =q^{-\h H_1\ot
H_2}\sum^{\infty}_{n=0} \frac{(1-q^2)^{n}}{[n]_{q}!} q^{-\h
n(n-1)}(q^{\h H_1}\tilde{X^{-}})^{n}\ot (q^{-\h
H_2}\tilde{X^{+}})^{n}, \label{Rg}\eeq which is very similar to
the universal ${\cal R}$ matrix of $U_{q}sl(2)$: \beq {\cal
R}_{sl}=q^{-\h H\ot
H}\sum^{\infty}_{n=0}\frac{(1-q^2)^{n}}{[n]_{q}!} q^{-\h
n(n-1)}(q^{\h H}\tilde{X^{-}})^{n}\ot (q^{-\h
H}\tilde{X^{+}})^{n}. \label{Rs} \eeq

In fact, the similarity is not an accident but a consequence of
QD. To see this, we first replace the generators $\tilde{X^{+}}$
and $\tilde{X^{-}}$ by $e$ and $f$ \cite{Tjin}: \beq
e=q^{-\frac{H_{2}}{2}}\tilde{X^{+}}\;,\;\;\;\; f=
q^{\frac{H_{1}}{2}},\tilde{X^{-}} \eeq and then the equation
(\ref{cmt})-(\ref{eee}) become
\[
[H_{1,2}\,,\,e]=2 e\;,\;\;\;[H_{1,2}\,,\,f]=-2 f\;,
\]
\beq [e\,,\,f]=\frac{q^{H_{1}}-q^{-H_{2}}}{q^{2}-1}\;, \eeq
\[
\dt(H_{1,2})=H_{1,2} \ot 1+1 \ot H_{1,2}\;,\;\dt (1) = 1 \ot 1\;,
\]
\beq \dt (e)=e \ot 1 + q^{-H_{2}} \ot e\;,\;{\dt (f)=1 \ot f + f
\ot q^{H_{1}}}. \eeq \beq S(H_{1,2})=-H_{1,2} \;,\;S(e)=-q^{H_{2}}
e \;,\;S(f)=-f q^{-H_{1}}\;,\;S(1)=1, \eeq \beq \vep (H_{1,2})
=\vep (e) =\vep (f) = 0\;,\;\vep (1) = 1. \eeq These equations
provide us the coefficients in the construction of quantum double.
Now, choosing the lower Borel subalgebra of $U_{q}gl(2)$
\[
U_{q}{\cal B}_{-}=span\{H_{1}^{n}f^{m}\}^{\infty}_{n,m=0}
\]
as the Hopf algebra $A$ in the quantum double construction, then
by applying the same method as Tjin did in \cite{Tjin}, we will
find that $A^{o}$ can be identified to the upper Borel subalgebra
\[
U_{q}{\cal B}_{+}=span\{H_{2}^{n}e^{m}\}^{\infty}_{n,m=0}
\]
and finally obtain the quantum double $D(A)$ as $U_{q}gl(2)$.

Note that in the case of $U_{q}sl(2)$, the dual element of $H$ can
only be identified to an element proportional to itself. However,
in the $U_{q}gl(2)$ case, the existence of commuting element $J$
makes it possible to identify the dual element of $H_{1}$ as
$H_{2}$, with
\[
H_{1}-H_{2}\propto J,
\]
thus establish the universal ${\cal R}$ matrix in equation
(\ref{Rg}).

The same multiparameter universal ${\cal R}$ matrix can also be
obtained in a different way. Denote ${\cal R}$ in (\ref{R}) as
${\cal R}(H_{1}\,,\,H_{2})$. Let $u=1$ and define
$q^{H'}=q^{H}t^{-J}$, we obtain a single-parameter $U_{q}gl(2)$
and universal ${\cal R}$ matrix denoted by ${\cal R}(H'\,,\,H')$:
\beq {\cal R}(H'\,,\,H')=q^{-\h H'\ot
H'}\sum^{\infty}_{n=0}\frac{(1-q^2)^{n}}{[n]_{q}!} q^{-\h
n(n-1)}(q^{\h H'}\tilde{X^{-}})^{n}\ot (q^{-\h
H'}\tilde{X^{+}})^{n}. \label{Rh'}\eeq According to the procedure
introduced by Reshetikhin \cite{Rs}: if we can find an element
$F=\sum_{i}f^{i}\ot f_{i}$ $\in$ $U_{q}gl(2)\ot U_{q}gl(2)$
satistying
\[
(\dt \ot id)F=F_{13}F_{23},\;\;\;\;\;(id \ot \dt)F=F_{13}F_{12},
\]
\beq F_{12}F_{13}F_{23}=F_{23}F_{13}F_{12},\;\;\;\;F_{12}F_{21}=1,
\eeq then we can build a multiparameter version of this QTHA and
thus obtain a multiparameter universal ${\cal R}$ matrix: \beq
{\cal R}^{(F)}=F^{-1}{\cal R}(H',H')F^{-1}. \eeq One can check
that \beq F=u^{-\frac{1}{4}(H\ot J-J\ot H)} \eeq can be used to do
this construction and the Hopf maps defined in
(\ref{cmt})-(\ref{eee}) and the universal ${\cal R}$ matrix in
(\ref{R}) will be recovered. Note that in the expression of ${\cal
R}_{0}$ (cf. equation (\ref{R0})), the exponent of the parameter
$u$ has an $antisymmetric$ form, which can be obtained from
Reshetikhin's transformation. On the other hand, the exponent of
the parameter $t$ has a symmetric form which comes from the third
formula of (\ref{cmt}), and $cannot$ be obtained from
Reshetikhin's transformation.
\section{The highest weight representations of $U_{q}gl(2)$}

For the representation theory, we only study the highest weight
representations \cite{Lee,GSX,KBM}. Let $\pi$ be the map from
$U_{q}gl(2)$ to a $m-$dimensional ( $m\geq 2$) representation:
\[
\pi(J)=\lambda {\bf 1},\;\;\;\pi(H)=\mu {\bf 1}+\sum^{m}_{i=1}(m
-2i+1)\,e_{ii},
\]
\beq \pi(\tilde{X^{+}})=
\sum^{m-1}_{i=1}a_{i}\,e_{i,i+1},\;\;\;\;\pi(\tilde{X^{-}})=
\sum^{m-1}_{i=1}b_{i}\,e_{i+1,i}, \eeq here $e_{ij}$ represents
the matrix basis ($(e_{ij})_{kl}=\delta_{ik}\delta_{jl}$) and
${\bf 1}$ denots the unit matrix. Our strategy is to find a proper
choice of parameters $\lambda$, $\mu$, $\{a_{i},
b_{i}\}^{m}_{i=0}$ such that they will give us the highest weight
representations of $U_{q}gl(2)$. Substituting these expressions to
(\ref{cmt}), we get \beq
a_{i}b_{i}=[i]_{q}\left(\frac{q^{\mu}q^{m-i}t^{-\lambda}-q^{-\mu}q^{i-m}
t^{\lambda}}{q-q^{-1}}\right),\;\;\;i=1,2,\ldots,m-1,\label{ab}
\eeq here $b_{i}$ does not have any prior relation to $a_{i}$.
Equation (\ref{ab}) naturally comes from the commutation relation
(\ref{cmt}) of $U_{q}gl(2)$. Let $t^{\lambda}=q^{\tau}$,
(\ref{ab}) can now be rewritten as:
\[
a_{i}b_{i}=[i]_{q}[\mu-\tau + m - i]_{q},\;\;\;\;i=1,2,\ldots,m-1.
\]
\ni For $i=m-1$, comparing with another expression (also obtained
from (\ref{cmt})):
\[
a_{m-1}b_{m-1}= -[\mu-\tau +1-m]_{q},
\]
and using the identities:
\[
[x]^{2}_{q}-[y]^{2}_{q}=[x-y]_{q}[x+y]_{q},
\]
\[
[x]_{q}[y]_{q}=[\frac{x+y}{2}]^{2}_{q}-[\frac{x-y}{2}]^{2}_{q},
\]
we find \beq [\mu-\tau]_{q}[m]_{q}=0. \eeq

This result thus gives us two kinds of the highest weight
representation:
\begin{itemize}
  \item{\bf Type a.} If $q^{2 (\mu-\tau)}=1$ or $q^{2\mu}t^{-2\lambda}=1$,
  then $q$ can be any complex number.
  \item{\bf Type b.} If $\mu, \tau$ or $q^{2\mu}t^{-2\lambda}$ are arbitrary
  complex numbers, then we must have the restriction $[m]_{q}=0$
  or $q^{2 m}$$=1$. In other words, $q$ must be restricted to roots of unity.
\end{itemize}

Now let's consider two simple examples. First the $m=2$ case:
\[
\pi(H)=\left(\bary{cc}{\mu +1}&0\\ 0&{\mu
-1}\eary\right),\;\;\;\;\:\pi(J)=\lambda\left(\bary{cc}1&0\\
0&1\eary\right),
\]
\beq \pi(\tilde{X^{+}})=\left(\bary{cc}0&a\\
0&0\eary\right),\;\;\pi(\tilde{X^{-}})=\left(\bary{cc}0&0\\
b&0\eary\right), \eeq \beq ab=\frac{q^{\mu +1}t^{-\lambda}-q^{-\mu
-1}t^{\lambda}}{q-q^{-1}}. \eeq

The $4\times 4$ matrix solutions $R$ of YBE can be obtained via
the representation $R=(\pi \ot\pi){\cal R}$:

\beq R=q^{-\h (\mu^2 -1)}t^{\lambda
\mu}\left(\bary{cccc}{q^{-1}(q^{-\mu}t^{\lambda})}&0&0&0\\
0&u^{\lambda}&0&0\\ 0&{(q^{-1}-q)ab}&u^{-\lambda}&0\\
0&0&0&{q^{-1}(q^{\mu}t^{-\lambda})}\eary\right). \eeq

\ni Let $q^{\mu}t^{-\lambda}= q^{-1}s$, $u^{\lambda}= \ga$ and
drop the factor $q^{-\h (\mu^2-1)}t^{\lambda \mu}$, we have

\beq R=\left(\bary{cccc}{s^{-1}}&0&0&0\\ 0&\ga&0&0\\
0&{s^{-1}-s}&{\ga^{-1}}&0\\ 0&0&0&{q^{-2}s}\eary\right). \eeq

According to previous discussion, this $R$ matrix in fact
represents two solutions, which are

\beq R_{(1)}=\left(\bary{cccc}s^{-1 }&0&0&0\\0&\gamma&0&0\\0&{s^
{-1}-s}&\gamma^{-1}&0\\0&0&0&s^{ -1}\eary\right),
\;\;R_{(2)}=\left(\bary{cccc }s^{-1}&0&0&0\\0&\gamma&0&0\\
0&{s^{-1}-s}&\gamma^{-1}&0\\ 0&0&0&{-s}\eary\right). \eeq

\ni When $q$ is generic, we must have $q^{-2}s^2=1$ which gives us
solution $R_{(1)}$. On the other hand, if $s$ is arbitrary, we
have $q^4=1$ which implies $q^2=-1$ ( $q^2=1$ is ruled out since
that will cause $ab\ra \infty$)and gives us solution $R_{(2)}$.
Next, we consider the $m=3$ case,
\[
\pi(H)=\left(\bary{ccc}{\mu +2}&0&0\\ 0&{\mu}&0\\ 0&0&{\mu
-2}\eary\right),\;\;\pi(J)=\lambda\left(\bary{ccc}1&0&0\\
0&1&0\\0&0&1\eary\right),
\]
\beq
\pi(\tilde{X^{+}})=\left(\bary{ccc}0&a_{1}&0\\0&0&a_{2}\\0&0&0\eary\right),\;\;
\pi(\tilde{X^{-}})=\left(\bary{ccc}0&0&0\\b_{1}&0&0\\0&b_{2}&0\eary\right),
\eeq

\[
a_{1}b_{1}=[\mu -\tau +2]_{q}=\left(\frac{q^{\mu
+2}t^{-\lambda}-q^{-\mu -2}t^{\lambda}}{q-q^{-1}}\right)
,\;\;t^{\lambda}=q^{\tau},
\]

\beq a_{2}b_{2}=[2]_{q}[\mu -\tau
+1]_{q}=(q+q^{-1})\left(\frac{q^{\mu +1}t^{-\lambda}-q^{-\mu
-1}t^{\lambda}}{q-q^{-1}}\right). \eeq

\ni Let $q^{-\mu}t^{\lambda}=q^2 s^{-2}$, $u^{\lambda}= \ga$ and
remove the factor $q^{-\h \mu^{2}}t^{\lambda \mu}$, we get

\beq R=\left(\bary{ccc}A_{1}&\large{0}&\large{0}\\
B_{1}&A_{2}&\large{0}\\ C&B_{2}&A_{3} \eary\right),\label{333}
\eeq
\\
where $A_{1}, A_{2}, A_{3}$, $B_{1}, B_{2}$ and $C$ are $3\times
3$ matrices:
\[
A_{1}=\left(\bary{ccc}{q^{2}s^{-4}}&0&0\\0&{q^{2}s^{-2}\ga}&0\\0&0&{q^{2}\ga^{2
}}\eary\right),\;\;\;A_{2}=\left(\bary{ccc}{q^{2}s^{-2}\ga^{-1}}&0&0\\0&1&0\\0
&0&{q^{-2}s^{2}\ga}\eary\right),
\]
\beq
A_{3}=\left(\bary{ccc}{q^{2}\ga^{-2}}&0&0\\0&{q^{-2}s^{2}\ga^{-1}}&0\\0&0&{q^{-
6}s^{4}}\eary\right), \eeq

\[
B_{1}=\left(\bary{ccc}0&{q^{2}(s^{-4}-1)}&0\\0&0&{ (1-q^{2})\ga
a_{2}b_{1}}\\0&0&0\eary\right),
\]
\beq B_{2}=\left(\bary{ccc}0&{(1-q^{2})\ga^{-1}
a_{1}b_{2}}&0\\0&0&{(1+q^{-2})(1-q^{-2}s^{4})}\\0&0&0\eary\right),
\eeq

\beq
C=\left(\bary{ccc}0&0&{(s^{-4}-1)(q^{2}-s^{4})}\\0&0&0\\0&0&0\eary\right).
\label{334} \eeq

This result also gives us two kinds of $R$ matrices. When
$(q/s)^4=1$, we have the type a. solution (the standard solution),
whereas in the situation $(q/s)^4\neq 1$, we have $q^6=1\ra
1+q^2+q^4=0$ which gives us type b. solution (the nonstandard
solution) respectively. Notice that the factors $a_{1}b_{2}$ and
$a_{2}b_{1}$ appearing in $B_1$ and $B_2$ cannot be uniquely
determined in terms of $q, \ga, s$ only, whereas their product
$(a_{1}b_{2}a_{2}b_{1})=(a_{1}b_{1}a_{2}b_{2})$ is unique. For a
general integer $m$, after removing the factor $q^{-\h
\mu^2}t^{\lambda \mu}$, and let \beq
q^{\mu}t^{-\lambda}=(q^{-1}s)^{m-1}\;,\;\;\;\;\;\;\;\;\;\;u^{\lambda}=\ga,
\eeq we have
\[
R=q^{\h
(m-1)^{2}}s^{-(m^{2}-1)}\sum^{m-1}_{n=0}\frac{(1-q^{2})^{n}}{\{n\}_{q^{2}}!}
q^{n}\sum^{m-n}_{i,j=1}q^{-2(i-1)(j-1)-n (i+j)}
\]
\beq s^{(m-1)(i+j+n)}\ga^{-(i-j)}
(a_{j}b_{i})\cdots(a_{j+n-1}b_{i+n-1})\,e_{i+n,i}\ot
e_{j,j+n},\label{sgab} \eeq where \bqn
a_{i}b_{i}&=&[i]_{q}\left(\frac{q^{\mu}t^{-\lambda}q^{m-i}-q^{-\mu}t^{\lambda}
q^{i-m}}{q-q^{-1}}\right)\non\\
&=&[i]_{q}\left(\frac{s^{m-1}q^{1-i}-s^{1-m}q^{i-1}}{q-q^{-1}}\label{aibi}\right)
\eqn and the identity \beq (q^{2 \mu}t^{-2 \lambda}-1)[m]_{q}
=\left((\frac{s}{q})^{2(m-1)}-1\right)[m]_{q}=0\label{rest} \eeq
is hold. Here we define: $(a_{j}b_{i})\cdots
(a_{j+n-1}b_{i+n-1})\equiv 1$ when $n=0$.

\section{Colored solutions of Yang-Baxter equation}

In order to obtain a colored solution of YBE via representation,
we have to prepare two representations of $U_{q}gl(2)$
\cite{GSX,KBM}: $\pi_1=\pi^{\mu,\lambda}$ and $\pi_2=\pi^{{\mu}',
{\lambda}'}$ acting on the former and later entries associated
with tensor product $\ot$ respectively. Then the colored solution
is given by \beq R(\mu,\lambda;{\mu}', {\lambda}')=(\pi_{1}\ot
\pi_{2}){\cal R} \eeq

Now let's calculate $R(\mu,\lambda;\mu^{'},\lambda^{'})$. For the
former entry associated with $\ot$, we have
\[
\pi_{1}(H)=\sum^{m}_{i=1}(\mu + m -2i +1)\,e_{i
i},\;\;\;\;\;\;\pi_{1}(J)=\lambda {\bf 1}=\lambda \sum^{m}_{i=1}\,
e_{ii},
\]
\[
\pi_{1}(\tilde{X^{-}})=\sum^{m-1}_{i=1}\,b_{i}\,e_{i+1,i},
\]
and for the later entry, we have
\[
\pi_{2}(H)=\sum^{m}_{i=1}({\mu}' + m -2i
+1)\,e_{ii},\;\;\;\;\;\;\pi_{2}(J)={\lambda}' {\bf
1}={\lambda}'\sum^{m}_{i=1}\, e_{ii},
\]
\[
\pi_{2}(\tilde{X^{+}})=\sum^{m-1}_{i=1}\,{a'}_{i}\,e_{i,i+1}.
\]
Here,
\[
R(\mu, \lambda \,;\, {\mu }',{\lambda }')=f(\mu
,\lambda\,;\,{\mu}',{\lambda }') \sum^{m-1}_{n=0}\frac{(1-q^2
)^{n}}{\{n\}_{q^2 }!} q^{n}(ss')^{\frac{n}{2}(m-1)}(\frac{\ga
}{{\ga}'})^{\frac{n}{2}}
\]
\[
\sum^{m-n}_{i,j=1}\,q^{-2(i-1)(j-1)-n(i+j)}((s')^{m-1}({\ga}')^{-1})^{i}
(s^{m-1}\ga )^{j}
\]
\beq ({a'}_{j}b_{i})\cdots ({a'}_{j+n-1}b_{i+n-1})\,e_{i+n,i}\ot
e_{j.j+n}, \label{solR}\eeq and $s\,,\,s'$, $\ga \,,\,{\ga }'$ are
defined by \beq (\frac{s}{q})^{m-1}=q^{\mu }t^{-\lambda
},\;\;\;(\frac{s'}{q})^{m-1}=q^{{\mu }'} t^{-{\lambda
}'},\;\;\;\ga =u^{\lambda },\;\;\;{\ga }'=u^{{\lambda }'}. \eeq
and the factor \beq f(\mu, \lambda\,;\,{\mu }',{\lambda }')=q^{-\h
\mu {\mu}'}t^{\h (\mu {\lambda }'+{\mu }'\lambda )}u^{\h (\mu
{\lambda}'-{\mu }'\lambda )} q^{\h (m-1)^{2}}(ss')^{-\h
(m^{2}-1)}(\frac{{\ga}'}{\ga })^{\h (m+1)} \eeq is irrelevant and
can be dropped.

As discussed in the last section, there are two different types of
solution: type a( q is generic ) and type b( q is a root of
unity). When $m=2$, let's compare our results with Hlavat\'{y}'s
solutions \cite{Hla} ( see also \cite{GX}):

\beq R_{1}(\lambda,\mu)=\phi
(\lambda,\mu)\left(\bary{cccc}1&0&0&0\\ 0&p^{+}(\lambda)&0&0\\
0&(1-k)\xi (\lambda)/\xi (\mu)&k/p^{+}(\mu)&0\\
0&0&0&p^{+}(\lambda)/p^{+}(\mu)\eary\right), \label{zxc} \eeq

\beq R_{2}(\lambda,\mu)=\phi (\lambda,\mu
)\left(\bary{cccc}1&0&0&0\\ 0&p^{+}(\lambda)&0&0\\
0&W(\lambda,\mu)&p^{-}(\mu)&0\\ 0&0&0&-p^{+}(\lambda )p^{-}(\mu)
\eary\right)\label{cxz}, \eeq where \beq
W(\lambda,\mu)=(1-p^{+}(\lambda)p^{-}(\lambda))\xi (\lambda)/
\xi(\mu)\label{lmw} \eeq with $\xi (\lambda)$ is an arbitrary
function.

\ni{\bf 1. } For type a:

\beq R_{a}=q^2 (\ga / {\ga}')\left(\bary{cccc}1&0&0&0\\
0&q\ga&0&0\\ 0&\pm (1-q^2 )(\ga / {\ga}')^{\h}&q/{\ga}'&0\\
0&0&0&\ga /{\ga}'\eary\right), \eeq

\ni which becomes $R_{1}$ when we define $p^{+}(\lambda)=q\ga$,
$p^{+}(\mu)=q{\ga}'$,$k=q^{2}$, $\xi (\lambda)/ \xi (\mu)$$=\pm
(\ga /{\ga}')^{\h}$.

\ni{\bf 2.} For type b:

\beq R_{b}=(ss')(\frac{\ga }{{\ga}'})\left(\bary{cccc}1&0&0&0\\
0&s{\ga}&0&0\\ 0&-2q(ss')^{\h}(\ga /{\ga}')^{\h} a'b&s'/{\ga}'&0\\
0&0&0&-ss'(\ga /{\ga}')\eary\right), \eeq

\ni here $q^{2}=-1$, $a'\,,\,b$ are arbitrary $C$ numbers. Let
$p^{+}(\lambda )=s \ga$, $p^{+}(\mu)=s'{\ga}'$, $p^{-}(\lambda)=s/
\ga$, $p^{-}(\mu)=s'/ {\ga}'$, we get the diagonal part of
$R_{2}$. Furthermore, rewriting $a'b=a'ab/a$, and using the
relation $ab=(s-s^{-1})$$/$$(q-q^{-1})$$=(q/2s)$$(1-s^2 )$ and
define \beq \frac{\xi (\lambda )}{\xi (\mu )}=\frac{[(\ga
/s)^{\h}/a]}{[({\ga}' /s')^{\h}/a']} \eeq we obtain
$W(\lambda\,,\;\mu)=-2q(ss')^{\h}(\ga  / {\ga}')^{\h} a'b$, which
leads to the-non standard solution $R_{2}$.

Another interesting application is to compare our solution with
that given in \cite{GSX}. Their universal ${\cal R}$ matrix (4.1)
is our $\bar{\cal R}$. The equivalence can be easily understood by
the replacements: \beq 2\hat{N}-\lambda_{1}\longrightarrow
H_{1},\;\;\;\; 2\hat{N}-\lambda_{2}\longrightarrow H_{2}, \eeq
\beq a^{\dagger}\cdot\alpha(\hat{N})\longrightarrow
\tilde{X^{+}},\;\;\; a\cdot\beta(\hat{N})\longrightarrow
\tilde{X^{-}}.\;\;\;\eeq The additional relation \beq
\alpha_{i}(\hat{N}-1)\cdot\beta_{i}(\hat{N})=[\lambda_{i}+1-\hat{N}]_{q}
\eeq appearing in \cite{GSX} is a consistency condition, just like
our equations (\ref{aibi}) and (\ref{rest}). Therefore, without
explicit calculation, the solutions obtained in \cite{GSX} are the
same as (\ref{solR}).

When comparing the solution (\ref{solR}) with those in
\cite{GLW,GOW,KBM,BH1}, one should be aware of the definitions and
conventions between ours and theirs( in particular, some authors
define our $RP$ or $PR$ as their $R$, $P$ represents the
permutation matrix ). Others even adopt different convention in
the definitions of $\dt$ and $S$. Therefore, one should first
properly choose a correct convention of $\{\dt, S\}$ and
definition of ${\cal R}$ or $R$.
\section{Concluding remarks}

We have studied the Hopf algebra structure and representation
theory of a multiparameter version of $U_{q}gl(2)$. We show that
the YBE can be solved directly in the QTHA framework, without
introducing additional tricks or doing any transformations. The
interesting feature of highest weight representation shows that
there exist two kinds of representations. A large class of Borel
type solutions $R$ can be obtained via the highest weight
representation, including standard and nonstandard colored
solutions. However, in this paper we have not yet discussed the
cyclic representation \cite{Jb1,GF,RA} of $U_{q}gl(2)$ for $q$
being a root of unity. We also have not explored what will happen
to the $U_{q}gl(2)$ algebra itself and its universal ${\cal R}$
matrix when $q$ is a root of unity \cite{GomS1,GomS2}. We leave
these discussions to another publication.

\ni{\bf Acknowledgement\/} We would like to thank M.H. Tu and D.H.
Lin for reading the manuscript. This work is partly supported by
grants NSC 87-M-2112-008-002 and NSC88-M-2112-008-009 from the
National Science Council (ROC).

\end{document}